\documentclass[12pt,a4paper,fleqn]{cernrep}
\usepackage{axodraw,cite,epsfig}

\usepackage{amsmath}
\usepackage{amssymb}
\usepackage{amsfonts}
\usepackage{amscd}
\usepackage{graphicx}

\begin{document}


\def\x{\chi}
\def\ti{\tilde}
\def\nt{\tilde \x^0}
\def\ch{\tilde \x^+}
\def\st{\tilde t}
\def\snu{\tilde\nu_\tau}
\def\stau{\tilde\tau}
\def\noi{\noindent}
\def\gravitino{\widetilde{G}} 
\def\Emiss{E_T^{\text{miss}}} 

\def\msg {m_{\tilde g}}
\def\msnu {m_{\snu}}
\newcommand{\mnt}[1]   {m_{\tilde\x^0_{#1}}}
\newcommand{\mch}[1]   {m_{\tilde\x^\pm_{#1}}}
\newcommand{\msf}[1]   {m_{\tilde f_{#1}}}
\newcommand{\mst}[1]   {m_{\tilde t_{#1}}}
\newcommand{\mstau}[1] {m_{\tilde\tau_{#1}}}
\newcommand{\msl}[1]   {m_{\tilde l_{#1}}}
\newcommand{\mse}[1]   {m_{\tilde e_{#1}}}

\def\micromegas {{\tt micrOMEGAs\,2.0}}
\def\softsusy {{\tt SOFTSUSY\,2.0.10}}
\def\spheno {{\tt SPheno\,2.2.3}}

\newcommand{\eq}[1]  {\mbox{(\ref{eq:#1})}}
\newcommand{\fig}[1] {Fig.~\ref{fig:#1}}
\newcommand{\Fig}[1] {Figure~\ref{fig:#1}}
\newcommand{\tab}[1] {Table~\ref{tab:#1}}
\newcommand{\Tab}[1] {Table~\ref{tab:#1}}
\newcommand{\figs}[1] {Figs.~\ref{fig:#1}}

\newcommand{\eVdist}{\kern-0.06667em}
\newcommand{\Ev}{{\text{e}\eVdist\text{V\/}}}     
\newcommand{\Kev}{{\text{ke}\eVdist\text{V\/}}}
\newcommand{\Mev}{{\text{Me}\eVdist\text{V\/}}}
\newcommand{\Gev}{{\text{Ge}\eVdist\text{V\/}}}
\newcommand{\Tev}{{\text{Te}\eVdist\text{V\/}}} 
\newcommand{\ev}{{\,\text{e}\eVdist\text{V\/}}}   
\newcommand{\kev}{{\,\text{ke}\eVdist\text{V\/}}}
\newcommand{\mev}{{\,\text{Me}\eVdist\text{V\/}}}
\newcommand{\gev}{{\,\text{Ge}\eVdist\text{V\/}}}
\newcommand{\tev}{{\,\text{Te}\eVdist\text{V\/}}}

\newcommand{\gsim}{\;\raisebox{-0.9ex}
           {$\textstyle\stackrel{\textstyle >}{\sim}$}\;}
\newcommand{\lsim}{\;\raisebox{-0.9ex}{$\textstyle\stackrel{\textstyle<}
           {\sim}$}\;}


\vspace*{-18mm}
\begin{flushright}
  CERN-PH-TH/2007-049\\
  DESY-07-029
\end{flushright}
\vspace*{2mm}

\begin{center}

{\Large\bf Collider signatures of gravitino dark matter\\[2mm] 
        with a sneutrino NLSP}\\[8mm]

{\large Laura Covi$^{\,1}$, Sabine Kraml$^{\,2}$}\\[4mm]

{\it $^{1}$\,Deutsches Elektronen-Synchrotron DESY, 
             D--22603 Hamburg, Germany\\
     $^{2}$\,CERN, CH--1211 Geneva 23, Switzerland}

\vspace*{4mm}

\begin{abstract}
\noindent
For gravitino dark matter with conserved R-parity and mass
in the GeV range, very strong constraints from Big Bang 
Nucleosynthesis exclude the popular NLSP candidates like 
neutralino and charged sleptons. 
In this letter we therefore draw attention
to the case of a sneutrino NLSP, that is naturally realised
in the context of gaugino mediation.
We find interesting collider signatures, characterised by
soft jets or leptons 
due to the small sneutrino--stau mass splitting. Moreover, 
the lightest neutralino can have visible decays into staus, 
and in some part of the parameter space also into selectrons 
and smuons. We also show the importance of coannihilation effects 
for the evaluation of the BBN constraints. 
\end{abstract}

\end{center}

\vspace*{2mm}


\section{Introduction}

If the gravitino is the lightest supersymmetric particle (LSP) and 
stable (with conserved R-parity) or sufficiently long-lived,
it is a good candidate for the Cold Dark Matter (CDM). 
At high temperatures, gravitinos are produced by thermal scatterings
even if they are not in thermal equilibrium. 
The resulting energy density is approximately given by 
\cite{Bolz:2000fu,Pradler:2006qh}
\begin{equation}
	\Omega_{3/2}^\text{th} h^2 \simeq
	0.27 \left( \frac{T_\text{R}}{10^{10}\gev} \right)
	\left( \frac{100\gev}{m_{3/2}} \right)
	\biggl( \frac{m_{\tilde g}}{1\tev} \biggr)^2 \;,
\end{equation}
where $m_{\tilde g}$ is the running gluino mass evaluated at low energy.
For a given $\msg$, the maximal possible reheating temperature 
$T_\text{R}$ is obtained for the heaviest allowed gravitino mass.
  
Gravitinos are also produced non-thermally via the decays of the 
next-to-lightest supersymmetric particle (NLSP), leading to 
\begin{equation}
	\Omega^\text{non-th}_{3/2}h^2 =
	\frac{m_{3/2}}{m_\text{NLSP}} \Omega_\text{NLSP}^\text{th}h^2 \;.
\end{equation}
Here $\Omega_\text{NLSP}^\text{th}h^2$ is the would-be relic density of 
the NLSP from thermal freeze-out if it did not decay. 
The total energy density of the gravitino LSP, 
$\Omega_{3/2} h^2=\Omega_{3/2}^\text{th} h^2 + 
\Omega^\text{non-th}_{3/2}h^2$,  
has to be equal or smaller than the cosmologically observed 
CDM density. 
In particular, if gravitinos should make up all the cold dark matter, 
$0.094\le \Omega_{3/2} h^2 \le 0.135$ \cite{Hamann:2006pf}.  
In general the right CDM abundance can be obtained
from both mechanisms for supersymmetric masses in the GeV--TeV 
region~\cite{Bolz:2000fu,Feng:2003xh}.

On the other hand, Big Bang Nucleosynthesis (BBN) severely constrains 
the nature, the lifetime and the freeze-out abundance of the NLSP. 
This is because the electromagnetic and hadronic energy released by 
the NLSP decays into the gravitino at comparatively late times 
($t>100$\,s) can alter the primordial abundances of light 
elements~\cite{Kawasaki:2004qu,Steffen:2006hw}. 
Moreover if the NLSP is charged, also bound state effects can
change heavily the rates of the nuclear reactions and modify the
BBN predictions~\cite{Pospelov:2006sc,Kohri:2006cn,Kaplinghat:2006qr}.
 
In fact, most NLSPs are incompatible with BBN, as long as their
lifetime is not shorter than $10^3$ s, i.e. the supersymmetric 
spectrum is very heavy, or their abundance is not strongly suppressed 
compared to that expected by thermal 
freeze-out, e.g. diluted by late entropy 
production~\cite{Hamaguchi:2007mp, Pradler:2006hh}.  
So in the minimal setting of simple freeze-out and masses for both 
gravitino and NLSP in the GeV range,
neutralino~\cite{Ellis:2003dn,Feng:2004mt, Roszkowski:2004jd,Cerdeno:2005eu,Jedamzik:2006xz} and
stau~\cite{Steffen:2006hw,Pospelov:2006sc,Cyburt:2006uv} NLSP are
incompatible with BBN.\footnote{Of course most of the
constraints are weakened or disappear for shorter NLSP lifetime,
i.e.\ lighter gravitino masses or larger NLSP masses. 
We recall that the NLSP lifetime is given approximately by
$\tau_{\rm NLSP} \simeq 
10^6\, {\rm s} \left(\frac{m_{3/2}}{10\,{\rm GeV}} \right)^2
\left(\frac{m_{\rm NLSP}}{100\,{\rm GeV}} \right)^{-5}$.} 
For completeness, let us mention that a stop
NLSP could be viable in some particular region of the supersymmetric
parameter space~\cite{Diaz-Cruz:2007fc}.  A sneutrino NLSP, on the
other hand, is neutral and decays mainly into gravitino and neutrino,
which are not electromagnetically or hadronically active. The BBN
bounds~\cite{Feng:2004zu,Kanzaki:2006hm} arising from the neutrino
interactions and the subdominant decay channel into quarks are much
weaker than those for a neutralino or charged slepton NLSP.  In this
study, we therefore consider a sneutrino NLSP as an interesting
alternative.
  
The paper is organised as follows. In Section~2 we briefly explain the
model of gaugino mediation. In Section~3 we discuss the sparticle
spectrum in this model, focusing in particular on the parameter range
which leads to a sneutrino NLSP. In Section~4 we evaluate the BBN
constraints on the sneutrino NLSP scenario, going beyond the
approximation used in~\cite{Kanzaki:2006hm}. In Section~5 we discuss
the signatures at LHC and ILC, and Section~6 finally contains our
conclusions.

\section{The model}

In general, in models of supersymmetry (SUSY) breaking with universal
scalar and gaugino masses, the right-chiral charged sleptons are
lighter than the left-chiral ones and the sneutrinos.  The reason is
that the running of $m_{\ti l_R}^2$ is dominated by $U(1)_Y$ D-term
contributions, while $m_{\ti l_L}^2$ receives $SU(2)_L$ and $U(1)_Y$
D-term corrections.  This picture changes, however, for non-universal
SUSY breaking parameters at the high scale, especially for
non-universal Higgs-mass parameters with $m_{H_1}^2-m_{H_2}^2>0$, see
e.g.~\cite{Ellis:2002iu}.

A particularly attractive realisation of non-universal boundary
conditions is the case of gaugino
mediation~\cite{Kaplan:1999ac,Chacko:1999mi}, where supersymmetry
breaking occurs on a four-dimensional brane within a
higher-dimensional theory. In such a setting, fields which live in
different places will naturally feel such breaking with different
strength.  Gauge and Higgs superfields living in the bulk couple
directly to the chiral superfield $S$ responsible for SUSY breaking,
which is localised on one of the four-dimensional branes.  The gaugino
and Higgs fields hence acquire soft SUSY-breaking masses at tree
level. Squarks and sleptons, on the
other hand, are confined to some other branes, without direct coupling
to $S$ and this yields no-scale boundary
conditions~\cite{Ellis:1984bm,Inoue:1991rk} for their masses.
We therefore have the following boundary conditions at the
compactification scale $M_C$ \cite{Chacko:1999mi}:
\begin{subequations}
\begin{align}
  &g_1 = g_2 = g_3 = g \simeq 1/\sqrt{2} \;, \\ &M_1 = M_2 = M_3 =
  m_{1/2} \;,\\ &m_0^2 = 0 \quad \text{for all squarks and sleptons},
  \\ &A_0 = 0 \\ &\mu, B\mu, m^2_{H_{1,2}} \neq 0 \;,
\end{align}
\label{eq:gauginomed}
\end{subequations} 
with GUT charge normalisation used for $g_1$.  The superparticle
spectrum is determined from these boundary conditions and the
renormalisation group equations.  The free parameters of the model are
hence $m_{1/2}$, $m^2_{H_1}$, $m^2_{H_2}$, $\tan\beta$, and the sign
of $\mu$; $|\mu|$ being determined by radiative electroweak symmetry
breaking.
 
The model favours moderate values of $\tan\beta$ between about 10 and
25.  The parameter ranges leading to a viable low-energy spectrum were
discussed in \cite{Buchmuller:2005ma,Evans:2006sj}, assuming $M_C =
M_\text{GUT}$. In \cite{Buchmuller:2006nx} it was shown that either
the lightest neutralino or the gravitino can be viable dark matter
candidates in this model.  In particular,
Ref.~\cite{Buchmuller:2006nx} discussed the possibility of a gravitino
LSP with a (tau-)sneutrino NLSP for $m_{1/2}=500$~GeV and
$\tan\beta=10$ and $20$. In this case, the sneutrino NLSP occurs for
$m^2_{H_2}\lsim 0.5$~TeV$^2$ and large values of $m^2_{H_1}$ of
roughly $2$--$3$~TeV$^2$. Ref.~\cite{Evans:2006sj} also discussed the
collider phenomenology of gaugino mediation, concentrating however on
the case of a neutralino LSP.

\section{Sparticle spectrum in gaugino mediation with a sneutrino NLSP}

We here investigate the SUSY spectrum in the gaugino-mediation model
in more detail. We assume that the gravitino is the LSP and
concentrate on scenarios with a sneutrino NLSP.  Following
\cite{Buchmuller:2005ma,Buchmuller:2006nx}, we take $m_t=172.5$~GeV,
$m_b(m_b)=4.25$~GeV and $\alpha_s^{\text{SM
}\overline{\text{MS}}}(M_Z)=0.1187$ as SM input parameters, and
consider $m_{3/2}=10$~GeV as lower bound for the gravitino mass (the
upper bound being given by the NLSP mass and the BBN
constraints). Moreover, we take $M_C = M_\text{GUT}$.  We use
\softsusy\ \cite{Allanach:2001kg} to compute the sparticle and Higgs
masses and mixing angles, and \micromegas\
\cite{Belanger:2001fz,Belanger:2004yn,Belanger:2006is} to compute
the primordial abundance of the NLSP.

\begin{figure}[t]\centering
\includegraphics[height=7cm,width=7cm]{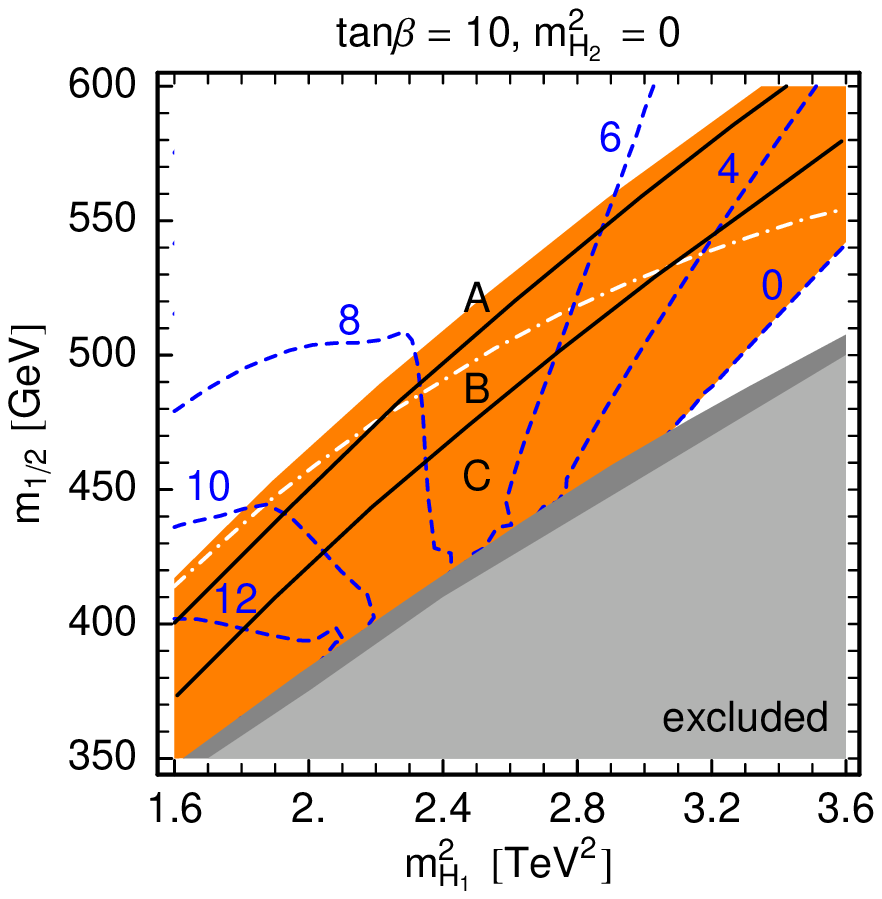}\quad
\includegraphics[height=7cm,width=7cm]{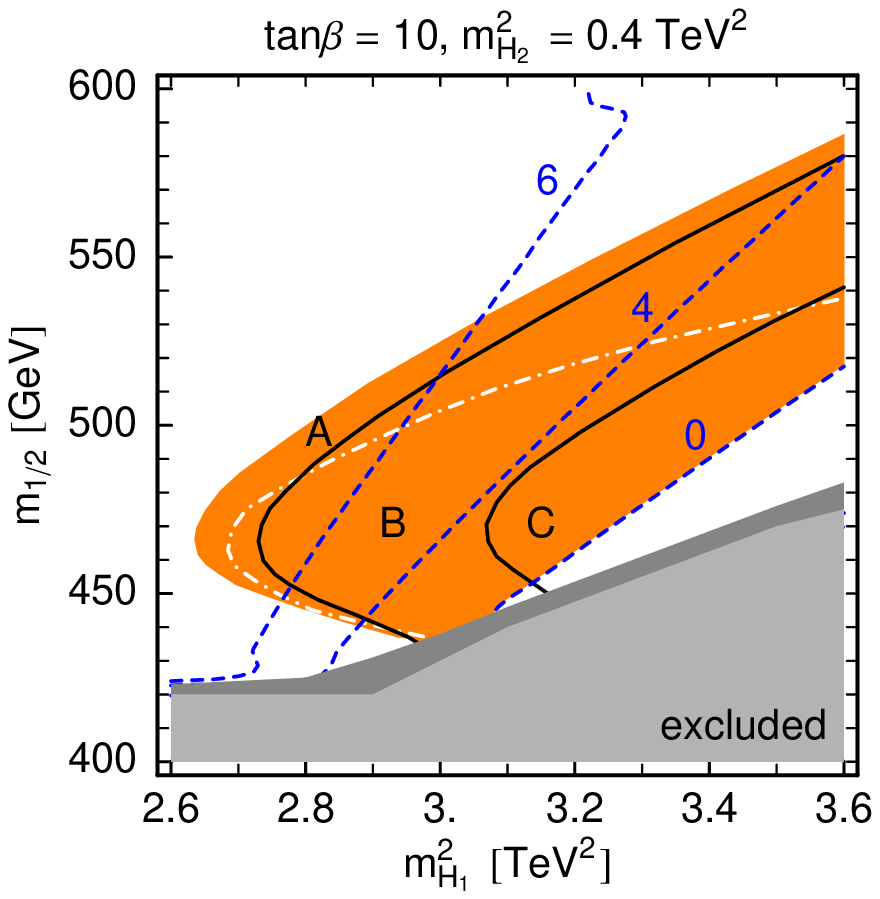}
\caption{Sneutrino NLSP regions (in orange) in the $m_{H_1}^2$ versus
$m_{1/2}$ plane for $\tan\beta=10$ and $m_{H_2}^2=0$ (left) and
$m_{H_2}^2=0.4$~TeV$^2$ (right).  The blue dashed lines show contours
of constant $\mstau{1}-\msnu$ in GeV.  The full black lines separate
subregions of different mass ordering:
$\msnu<\mnt{1}<\mstau{1}<\mse{L}$ in A,
$\msnu<\mstau{1}<\mnt{1}<\mse{L}$ in B, and
$\msnu<\mstau{1}<\mse{L}<\mnt{1}$ in C.  Below the white dash-dotted
line, the BBN bounds are satisfied for any gravitino mass, i.e.\
$m_{\tilde\nu} Y_{\tilde\nu} \leq 3 \times 10^{-11}$ GeV, as discussed
in the text.  In the light grey regions, no viable spectrum is
obtained, while in the narrow medium grey strips, $\mstau{1}<90$~GeV.}
\label{fig:snuLSP_region_tb10}
\end{figure}

\Fig{snuLSP_region_tb10} shows the sneutrino NLSP region in the
$m_{H_1}^2$ versus $m_{1/2}$ plane for $\tan\beta=10$ and two values
of $m_{H_2}^2$, $m_{H_2}^2=0$ and $0.4$~TeV$^2$.  Also shown are
contours of constant $\mstau{1}-\msnu$ in GeV: since $\mstau{L}$ and
$\msnu$ are driven by the same SUSY-breaking parameter $M_{\ti L_3}$,
the mass difference between the $\snu$ and the $\stau_1$ is always
small.  The mass of the $\snu$ NLSP goes up to about 250 (230)~GeV for
$m_{H_2}^2=0$ ($0.4$~TeV$^2$) and $m_{1/2}=600$~GeV in
\fig{snuLSP_region_tb10}.  Comparing with Fig.~4 of
\cite{Kanzaki:2006hm}, one might conclude that the $\snu$ NLSP region
of \fig{snuLSP_region_tb10} is in good agreement with BBN; this is
discussed in more detail in the next section.  For fixed $m_{1/2}$,
$\msnu$ decreases with increasing $m_{H_1}^2$, and so do $\mstau{1}$
and $\mse{L}\simeq m_{\ti\mu_L}$, while $\mnt{1}$ remains
constant. One therefore finds the mass orderings%
\footnote{Since selectrons and smuons are practically degenerate, in
the following $\ti e$ implicitly means selectrons and smuons.}
$\msnu<\mnt{1}<\mstau{1}<\mse{L}$, $\msnu<\mstau{1}<\mnt{1}<\mse{L}$
and $\msnu<\mstau{1}<\mse{L}<\mnt{1}$ within the sneutrino NLSP
region.  These are labelled A, B, and C, respectively,
in~\fig{snuLSP_region_tb10}.

\begin{figure}[t]\centering
\includegraphics[height=7cm,width=7cm]{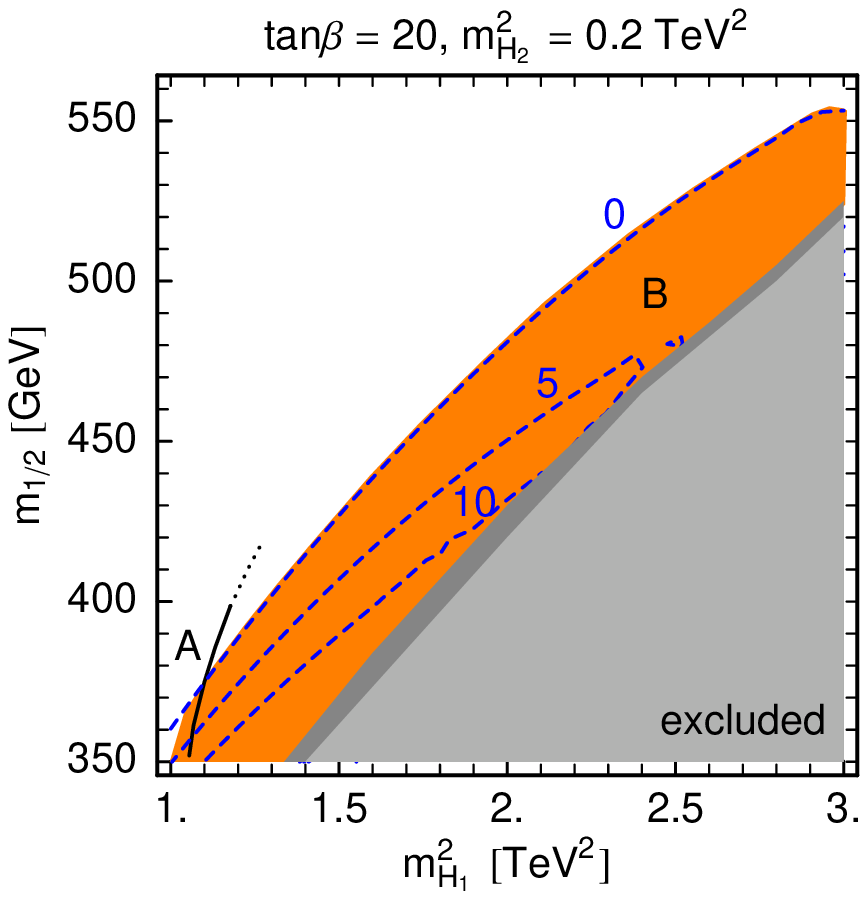}
\includegraphics[height=7cm,width=7cm]{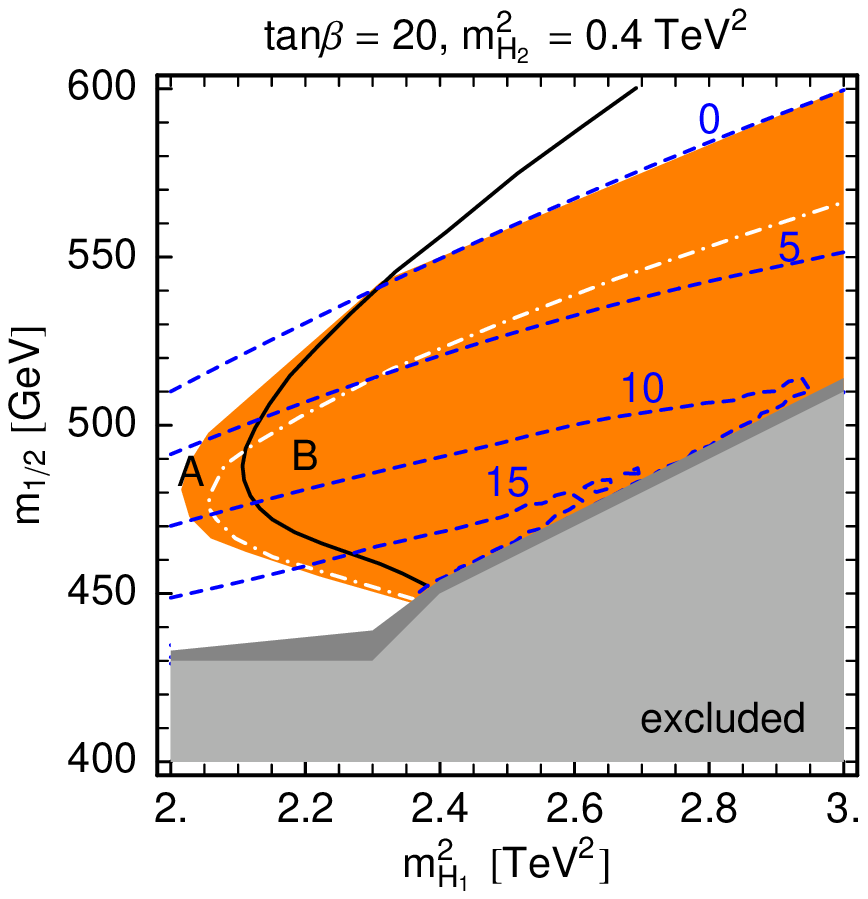}
\caption{Same as \fig{snuLSP_region_tb10} but for $\tan\beta=20$ and
$m_{H_2}^2=0.2$~TeV$^2$ (left) and $m_{H_2}^2=0.4$~TeV$^2$ (right).
BBN bounds play no role in the left-hand panel.}
\label{fig:snuLSP_region_tb20}
\end{figure}

The case of $\tan\beta =20$ is shown in \fig{snuLSP_region_tb20} for
$m_{H_2}^2=0.2$ and $0.4$~TeV$^2$.  Analogous arguments as above
apply. Note, however, that here the $\ti e_L^{}$ does not become
lighter than the $\nt_1$.  Moreover, the $\snu$--$\stau_1$ mass
difference shows a different behaviour as compared to $\tan\beta =10$:
At $\tan\beta=10$ and small $m_{H_1}^2$, $\msnu < \mstau{1}$ with the
mass difference becoming smaller as $m_{H_1}^2$ increases. At
$\tan\beta=20$, the $\stau_1$ is first lighter than the $\snu$; with
increasing $m_{H_1}^2$, $\msnu$ decreases faster than $\mstau{1}$,
eventually leading to $\msnu < \mstau{1}$. This is why the contour of
$\mstau{1}-\msnu=0$ is on the upper-left edge of the $\snu$ NLSP
region in \fig{snuLSP_region_tb20}, while it is on the lower-right
edge in \fig{snuLSP_region_tb10}.

A comment is in order concerning the LEP limit on the light Higgs
mass.  Demanding $m_{h^0}\ge 114.5$~GeV would constrain $m_{1/2}$ to
$m_{1/2}\gsim 500$ ($440$)~GeV in \fig{snuLSP_region_tb10}
(\ref{fig:snuLSP_region_tb20}).  However, there is still a 2--3~GeV
uncertainty in the evaluation of $m_{h^0}$. If this is taken into
account, the full parameter range considered is allowed.

\section{Sneutrino abundance and BBN constraints}

Even if the sneutrino is neutral and decays mainly into weakly
interacting particles, still BBN constraints arise from the subleading
decay channels.  According to~\cite{Kanzaki:2006hm}, Figure~4, such
bounds are satisfied for light sneutrinos with masses below 300 GeV,
because the branching ratios into quarks via virtual $Z,\,W$ are
rather small.  This conclusion was obtained through an estimate of the
sneutrino freeze-out abundance of
\begin{equation}
   Y_{\tilde\nu} \simeq 2 \times 10^{-14}
   \left(\frac{m_{\tilde\nu}}{100\,\mbox{GeV}} \right)\; .
\end{equation} 
In our case though, due to the close spacing between the different
masses, co-annihilation effects~\cite{Ellis:2002iu} become important,
making this estimate unreliable.  Here note that co-annihilation
effects can both decrease or increase the particle yield.  The latter
can occur if the co-annihilation cross section is small, due to the
presence in the thermal bath of the slightly heavier states that can
decay into the NLSP~\cite{Asaka:2000zh}.  We therefore use
\micromegas\ \cite{Belanger:2001fz,Belanger:2004yn,Belanger:2006is} to
compute $Y_{\tilde\nu}$ numerically without approximation, and obtain
that in our region of the parameter space the sneutrino abundance
\begin{equation}
   m_{\tilde\nu} Y_{\tilde\nu} = 3.63 \times 10^{-9} \mbox{GeV} \;
   \Omega_{\tilde\nu}^\text{th}\; h^2
\end{equation}
can be as large as $10^{-10}$~GeV.  This value violates the general
bounds given in~\cite{Kanzaki:2006hm} for a gravitino mass in the
range $2$--$50$ GeV.  The limit for a gravitino with a mass of about
$10$~GeV is in fact $m_{\tilde\nu} Y_{\tilde\nu} < 3 \times 10^{-11}$
GeV, which is shown as dash-dotted line in \figs{snuLSP_region_tb10}
and \ref{fig:snuLSP_region_tb20}.  For a gravitino mass of 50 GeV or
larger, or for a sneutrino decay branching ratio into hadrons 
substantially smaller than $10^{-3}$, 
this BBN bound becomes much weaker and disappears in our
parameter region.  We will consider in the following benchmark points
where the BBN constraints are satisfied.

Last but not least, since $\Omega_\text{NLSP}^\text{th}h^2$ is very
small, typically ${\cal O}(10^{-3})$, throughout the $\snu$ NLSP
region, $\Omega^\text{non-th}_{3/2}h^2$ is negligible and almost all
the gravitino dark matter has to be produced thermally.  Requiring
$\Omega_{3/2} h^2\simeq 0.1$ leads to $T_R\sim 10^8$--$10^9$~GeV for
$\msg\sim 1$~TeV and $m_{\gravitino}$ in the range of $10$--$100$~GeV.

\section{Collider signatures}

The collider signatures are characterised by the small
$\snu$--$\stau_1$ mass difference. As mentioned, we can have the cases
$\mstau{1}>\mnt{1}>\msnu$ (region A) or $\mnt{1}>\mstau{1}>\msnu$
(region B).
In the former the $\nt_1$ decays via $\nt_1\to\nu\snu$,
while in the latter it can also decay directly into the visible
channel $\nt_1\to\tau\stau_1$.  If also the $\ti e_L$ is lighter than
the $\nt_1$ (region C), $\nt_1\to e^\pm\ti e^\mp_L$ is possible in
addition.  The NLSP decay into the gravitino, $\snu\to\nu\gravitino$,
is of course invisible, regardless of the $\snu$ lifetime. On the
other hand, even if such a decay is impossible to detect, it is clear
that the sneutrino cannot be stable and the dominant DM component,
since it has been already excluded by direct
searches~\cite{Falk:1994es}.

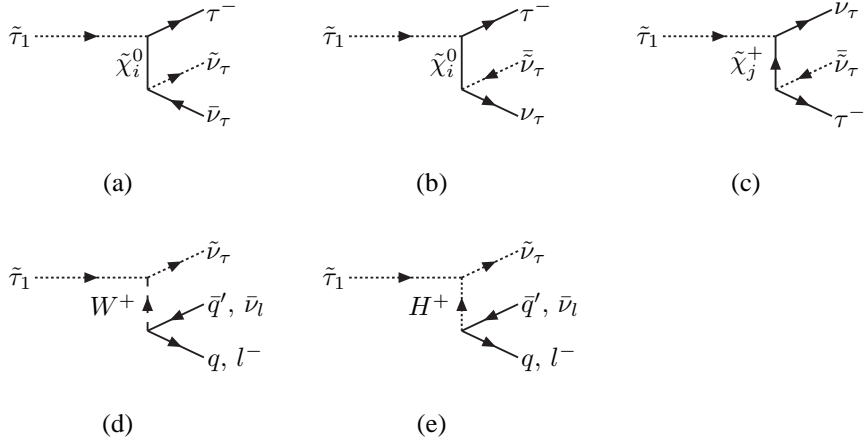
\begin{figure}\centerline {%
\unitlength=1.0 pt \SetScale{1.0} \SetWidth{0.7} 
\footnotesize 
\begin{tabular}{l}
%
\begin{picture}(95,79)(0,0)
\Text(15.0,60.0)[r]{$\tilde{\tau}_1$}
\DashArrowLine(16.0,60.0)(58.0,60.0){1.0}
\Text(80.0,70.0)[l]{$\tau^-$} \ArrowLine(58.0,60.0)(79.0,70.0)
\Text(57.0,50.0)[r]{$\tilde\chi^0_i$} \Line(58.0,60.0)(58.0,40.0)
\Text(80.0,50.0)[l]{$\tilde{\nu}_\tau$}
\DashArrowLine(58.0,40.0)(79.0,50.0){1.0}
\Text(80.0,30.0)[l]{$\bar{\nu}_\tau$} \ArrowLine(79.0,30.0)(58.0,40.0)
\Text(47,0)[b] {(a)}
\end{picture} 
\qquad
\begin{picture}(95,79)(0,0)
\Text(15.0,60.0)[r]{$\tilde{\tau}_1$}
\DashArrowLine(16.0,60.0)(58.0,60.0){1.0}
\Text(80.0,70.0)[l]{$\tau^-$} \ArrowLine(58.0,60.0)(79.0,70.0)
\Text(57.0,50.0)[r]{$\tilde\chi^0_i$} \Line(58.0,60.0)(58.0,40.0)
\Text(80.0,50.0)[l]{$\bar{\tilde{\nu}}_\tau$}
\DashArrowLine(79.0,50.0)(58.0,40.0){1.0}
\Text(80.0,30.0)[l]{$\nu_\tau$} \ArrowLine(58.0,40.0)(79.0,30.0)
\Text(47,0)[b] {(b)}
\end{picture}
\qquad
\begin{picture}(95,79)(0,0)
\Text(15.0,60.0)[r]{$\tilde{\tau}_1$}
\DashArrowLine(16.0,60.0)(58.0,60.0){1.0}
\Text(80.0,70.0)[l]{$\nu_\tau$} \ArrowLine(58.0,60.0)(79.0,70.0)
\Text(54.0,50.0)[r]{$\tilde\chi^+_j$} \ArrowLine(58.0,40.0)(58.0,60.0)
\Text(80.0,50.0)[l]{$\bar{\tilde{\nu}}_\tau$}
\DashArrowLine(79.0,50.0)(58.0,40.0){1.0}
\Text(80.0,30.0)[l]{$\tau^-$} \ArrowLine(58.0,40.0)(79.0,30.0)
\Text(47,0)[b] {(c)}
\end{picture} \\[3mm]
%
\begin{picture}(95,79)(0,0)
\Text(15.0,60.0)[r]{$\tilde{\tau}_1$}
\DashArrowLine(16.0,60.0)(58.0,60.0){1.0}
\Text(80.0,70.0)[l]{$\tilde{\nu}_\tau$}
\DashArrowLine(58.0,60.0)(79.0,70.0){1.0} \Text(54.0,50.0)[r]{$W^+$}
\DashArrowLine(58.0,40.0)(58.0,60.0){3.0} \Text(80.5,50.0)[l]{$\bar
q',\,\bar\nu_l$} \ArrowLine(79.0,50.0)(58.0,40.0)
\Text(80.5,30.0)[l]{$q,\,l^-$} \ArrowLine(58.0,40.0)(79.0,30.0)
\Text(47,0)[b] {(d)}
\end{picture}  
\qquad
\begin{picture}(95,79)(0,0)
\Text(15.0,60.0)[r]{$\tilde{\tau}_1$}
\DashArrowLine(16.0,60.0)(58.0,60.0){1.0}
\Text(80.0,70.0)[l]{$\tilde{\nu}_\tau$}
\DashArrowLine(58.0,60.0)(79.0,70.0){1.0} \Text(54.0,50.0)[r]{$H^+$}
\DashArrowLine(58.0,40.0)(58.0,60.0){1.0} \Text(80.5,50.0)[l]{$\bar
q',\,\bar\nu_l$} \ArrowLine(79.0,50.0)(58.0,40.0)
\Text(80.5,30.0)[l]{$q,\,l^-$} \ArrowLine(58.0,40.0)(79.0,30.0)
\Text(47,0)[b] {(e)}
\end{picture} 
\end{tabular}
}
\caption{Feynman diagrams for stau three-body decays into a sneutrino
LSP ($i=1...4$, $j=1,2$). The dominant contribution comes from the $W$
exchange of diagram (d). \label{fig:feyngraphs}}
\end{figure}

The $\stau_1$ can decay into $\tau\nt_1$ if
$\mstau{1}>\mnt{1}+m_\tau$; its 2-body decays into the NLSP,
$\stau_1^\pm\to W^\pm\snu$ or $H^\pm\snu$, are however kinematically
forbidden due to the small mass splittings.  For
$\mstau{1}<\mnt{1}+m_\tau$, the $\stau_1$ hence only has 3-body decays
leading to $f\!\bar f'$ plus missing energy as shown in
\fig{feyngraphs}.  The dominant contribution comes from the diagram
with the virtual $W$ boson.  The resulting $\stau_1$ lifetime in this
channel is approximately given by
\begin{equation}
  \Gamma_{\tilde\tau}^{-1} \simeq \frac{2 (2\pi)^3}{3 G_F^2
  m_{\tilde\tau}^5}\; F^{-1}
  \left(\frac{m_{\tilde\nu}^2}{m_{\tilde\tau}^2} \right) = 0.8\times
  10^{-16} \mbox{s} \; \left(\frac{m_{\tilde\tau}}{100\,\mbox{GeV}}
  \right)^{-5} \left(\frac{F \left(m_{\tilde\nu}^2/m_{\tilde\tau}^2
  \right)}{F(0.9)} \right)^{-1}
\end{equation}
where, after neglecting the $W$ momentum and the SM particle masses,
we have
\begin{equation}
F \left(a \right) = \int_{2\sqrt{a}}^{1+a} dx (x^2 - 4 a)^{3/2}\; .
\end{equation}
So for $\mstau{1}-\msnu\sim 5$--$10$~GeV the lifetime is of the order
of $10^{-16}$--$ 10^{-18}$~s; a displaced vertex is only obtained if
the $\snu$ and the $\stau_1$ are quasi-degenerate.

%

In the parameter range we consider, squarks and gluinos have masses of
about 1 TeV, leading to large SUSY cross sections at the LHC.  Since
$m_0=0$, the gluino is always the heaviest sparticle and decays into
$q\ti q$. Moreover, $\mse{L}<\mse{R}$ and the left-chiral sleptons can
be light enough to be produced in cascade decays.\footnote{This is in
sharp contrast to the CMSSM/mSUGRA case, where $\mse{L}>\mse{R}$, and
typically only the right sleptons appear in the cascades.}  In the
following, we discuss these cascade decays in more detail. If the
$\nt_1$ is mainly a bino (which is the case for zero or small
$m_{H_2}^2$), right-chiral squarks dominantly decay into $q\nt_1$.  If
$\mstau{1}+m_\tau>\mnt{1}>\msnu$, this looks just like the
neutralino-LSP case.  If, however, $\mnt{1}>\mstau{1}+m_\tau>\msnu$,
then the $\nt_1$ can decay further into
$\nt_1\to\tau^\pm\stau_1^\mp\to \tau^\pm f\!\bar f'\snu$.  Here note
that the $f\!\bar f'=(q\bar q',\,l\nu_l^{})$ will be quite soft.  The
left-chiral squarks can have more complicated cascade decays.  If
$\mnt{1}\gsim\mstau{1}$, these are generically given by the
conventional cascade decays into the $\nt_1$ as in the CMSSM, partly
supplemented by $\nt_1\to\tau^\pm\stau_1^\mp\to \tau^\pm f\!\bar
f'\snu$.  The resulting signatures are missing energy plus jets plus
\mbox{(single or di-)} leptons PLUS an additional tau, plus additional
soft leptons or jets if they can be detected. Examples for such
cascades are depicted in \fig{LHCcascades}.  The benchmark point no.~2
of \cite{Buchmuller:2005ma} with $m_{1/2}=500$~GeV, $\tan\beta=10$,
$m_{H_1}^2=2.7$~TeV$^2$, $m_{H_2}^2=0$ is an illustrative case. The
mass spectrum and the most important branching ratios for this point
are given in \tab{buchmuellerP2}.  The 2-body decays were computed
with {\tt SDECAY} \cite{Muhlleitner:2003vg}, and the 3-body decay with
{\tt CALCHEP} \cite{Pukhov:2004ca}.  The resulting ratios for the
decay chains of \fig{LHCcascades} are (a)~33\%, (b)~6\%, (c)~6.4\%,
(d)~3.3\%, (e)~7\%.  The sparticle masses can be determined from these
cascades through the standard method of invariant-mass distributions
of the SM decay
products~\cite{Hinchliffe:1996iu,Bachacou:1999zb,Allanach:2000kt,
Lester:2001zx,Miller:2005zp}; see also~\cite{ATLAS:1999fr,CMS:PTDR}
and references therein.  The correct interpretation of the scenario
is, however, more involved than in the conventional CMSSM case, and
care is needed in order not to falsely conclude to have found SUSY
with a neutralino LSP.  Notice also that the chain (e) as well as the
$\stau_1\to W^*\snu$ decays may fake lepton number violation.

%
\begin{figure}[p]\centering
\unitlength=1.0pt \SetScale{1.0} 
\SetWidth{1.0} \footnotesize 
\begin{picture}(240,80)
\DashArrowLine(0,10)(40,10){3.0} 
\ArrowLine(40,10)(80,10) 
\ArrowLine(40,10)(40,40)
\DashArrowLine(80,10)(120,10){3.0} 
\ArrowLine(80,10)(80,40)
\DashArrowLine(120,10)(160,10){3.0} 
\Line(120,10)(124,40)
\Line(120,10)(126,40) \Text(-3,11)[r]{$\tilde q_R^{}$}
\Text(60,3)[t]{$\tilde\chi^0_1$} \Text(100,3)[t]{$\tilde\tau_1^\mp$}
\Text(166,11)[l]{$\snu$} \Text(40,45)[b]{$q$}
\Text(80,45)[b]{$\tau^\pm$} \Text(126,45)[b]{$f\!\bar f'$}
\Text(-30,45)[r] {(a)}
\end{picture} 
\begin{picture}(240,80)
\DashArrowLine(0,10)(40,10){3.0} 
\ArrowLine(40,10)(80,10) 
\ArrowLine(40,10)(40,40)
\DashArrowLine(80,10)(120,10){3.0} 
\ArrowLine(80,10)(80,40)
\DashArrowLine(120,10)(160,10){3.0} 
\Line(120,10)(124,40)
\Line(120,10)(126,40) \Text(-3,11)[r]{$\tilde q_L^{}$}
\Text(60,3)[t]{$\tilde\chi^0_2$} \Text(100,3)[t]{$\tilde\tau_1^\mp$}
\Text(166,11)[l]{$\snu$} \Text(40,45)[b]{$q$}
\Text(80,45)[b]{$\tau^\pm$} \Text(126,45)[b]{$f\!\bar f'$}
\Text(-30,45)[r] {(b)}
\end{picture} 
\begin{picture}(240,80)
\DashArrowLine(0,10)(40,10){3.0} 
\ArrowLine(40,10)(40,40)
\ArrowLine(40,10)(80,10) 
\ArrowLine(80,10)(80,40)
\DashArrowLine(80,10)(120,10){3.0} 
\ArrowLine(120,10)(120,40) \ArrowLine(120,10)(160,10)
\ArrowLine(160,10)(160,40) \DashArrowLine(160,10)(200,10){3.0}
\Text(-3,11)[r]{$\tilde q_L^{}$} \Text(60,3)[t]{$\tilde\chi^0_2$}
\Text(100,3)[t]{$\tilde l^\mp_L$} \Text(140,3)[t]{$\tilde\chi^0_1$}
\Text(206,11)[l]{$\snu$} \Text(40,45)[b]{$q$} \Text(80,45)[b]{$l^\pm$}
\Text(120,45)[b]{$l^\mp$} \Text(160,45)[b]{$\nu_\tau$}
\Text(-30,45)[r] {(c)}
\end{picture} 
\begin{picture}(240,80)
\DashArrowLine(0,10)(40,10){3.0} 
\ArrowLine(40,10)(40,40)
\ArrowLine(40,10)(80,10) 
\ArrowLine(80,10)(80,40)
\DashArrowLine(80,10)(120,10){3.0} 
\ArrowLine(120,10)(120,40) \ArrowLine(120,10)(160,10)
\ArrowLine(160,10)(160,40) \DashArrowLine(160,10)(200,10){3.0}
\Line(200,10)(204,40) \Line(200,10)(206,40)
\DashArrowLine(200,10)(240,10){3.0} \Text(-3,11)[r]{$\tilde q_L^{}$}
\Text(60,3)[t]{$\tilde\chi^0_2$} \Text(100,3)[t]{$\tilde l^\mp_L$}
\Text(140,3)[t]{$\tilde\chi^0_1$} \Text(180,3)[t]{$\tilde\tau_1^\mp$}
\Text(246,11)[l]{$\snu$} \Text(40,45)[b]{$q$} \Text(80,45)[b]{$l^\pm$}
\Text(120,45)[b]{$l^\mp$} \Text(160,45)[b]{$\tau^\pm$}
\Text(207,45)[b]{$f\!\bar f'$} \Text(-30,45)[r] {(d)}
\end{picture} 
\begin{picture}(240,80)
\DashArrowLine(0,10)(40,10){3.0} \ArrowLine(40,10)(40,40)
\ArrowLine(40,10)(80,10) \ArrowLine(80,10)(80,40)
\DashArrowLine(80,10)(120,10){3.0} \ArrowLine(120,10)(120,40)
\ArrowLine(120,10)(160,10) \ArrowLine(160,10)(160,40)
\DashArrowLine(160,10)(200,10){3.0} \Line(200,10)(204,40)
\Line(200,10)(206,40) \DashArrowLine(200,10)(240,10){3.0}
\Text(-3,11)[r]{$\tilde q_L^{}$} \Text(60,3)[t]{$\tilde\chi^\pm_1$}
\Text(100,3)[t]{$\tilde l^\pm_L$} \Text(140,3)[t]{$\tilde\chi^0_1$}
\Text(180,3)[t]{$\tilde\tau_1^\mp$} \Text(246,11)[l]{$\snu$}
\Text(40,45)[b]{$q'$} \Text(80,45)[b]{$\nu_l$}
\Text(120,45)[b]{$l^\pm$} \Text(160,45)[b]{$\tau^\pm$}
\Text(207,45)[b]{$f\!\bar f'$} \Text(-30,45)[r] {(e)}
\end{picture} 
\caption{Examples of squark cascade decays in gaugino mediation with a
sneutrino NLSP; $l=(e,\,\mu)$.
\label{fig:LHCcascades}}
\end{figure}
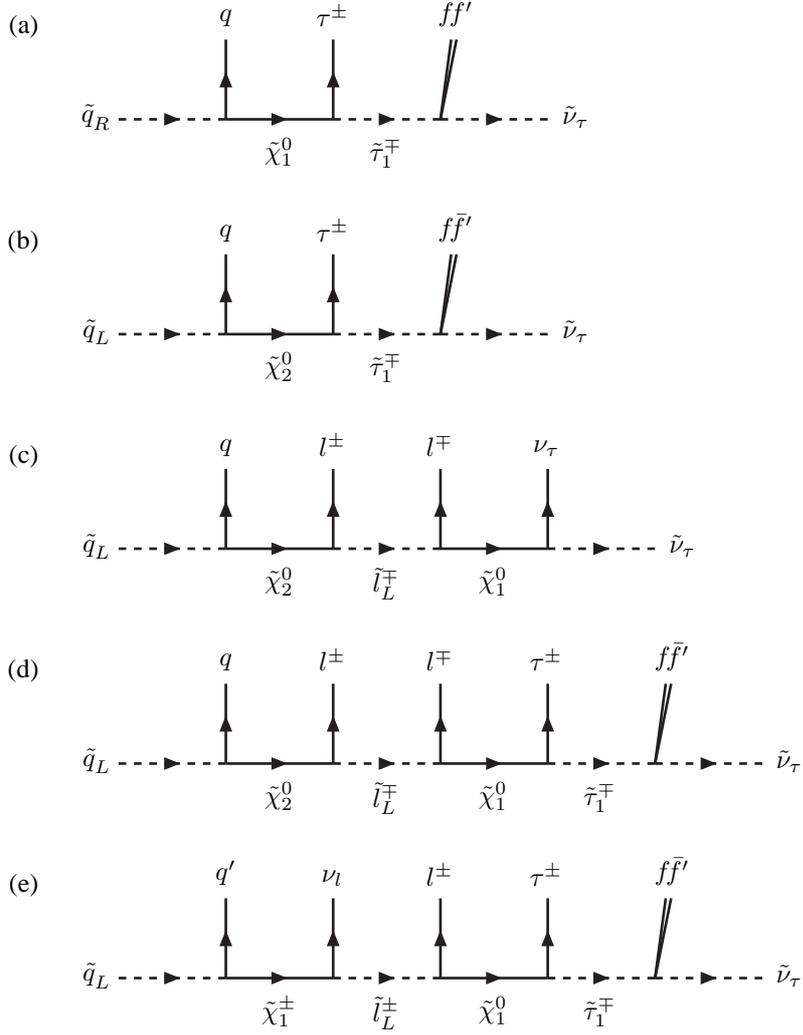

%
\begin{figure}[p]\centering
\unitlength=1.0pt \SetScale{1.0} 
\SetWidth{1.0} \footnotesize 
\begin{picture}(160,80)
\DashArrowLine(0,10)(40,10){3.0} 
\ArrowLine(40,10)(80,10) 
\ArrowLine(40,10)(40,40)
\DashArrowLine(80,10)(120,10){3.0} 
\ArrowLine(80,10)(80,40)
\DashArrowLine(120,10)(160,10){3.0} 
\Line(120,10)(124,40)
\Line(120,10)(126,40) \Text(-3,11)[r]{$\tilde q_{R(L)}^{}$}
\Text(60,3)[t]{$\tilde\chi^0_{1(2)}$} \Text(100,3)[t]{$\tilde
l_L^\mp$} \Text(166,11)[l]{$\snu$} \Text(40,45)[b]{$q$}
\Text(80,45)[b]{$l^\pm$} \Text(126,44)[b]{$l^\mp\nu_l$}
\Text(-20,45)[r] {(a)}
\end{picture} \\ 
\begin{picture}(160,80)
\DashArrowLine(0,10)(40,10){3.0} 
\ArrowLine(40,10)(80,10) 
\ArrowLine(40,10)(40,40)
\DashArrowLine(80,10)(120,10){3.0} 
\ArrowLine(80,10)(80,40)
\DashArrowLine(120,10)(160,10){3.0} 
\Line(120,10)(124,40)
\Line(120,10)(126,40) \Text(-3,11)[r]{$\tilde q_{L}^{}$}
\Text(60,3)[t]{$\tilde\chi^\pm_{1}$} \Text(100,3)[t]{$\tilde\nu_l$}
\Text(166,11)[l]{$\snu$} \Text(40,45)[b]{$q'$}
\Text(80,45)[b]{$l^\pm$} \Text(126,44.5)[b]{$\nu_l\nu_\tau$}
\Text(-20,45)[r] {(b)}
\end{picture} 
\caption{Examples of squark cascade decays
for the case $\mnt{1}>m_{\ti l_L}$ [in addition to
\fig{LHCcascades}(a,b)].
\label{fig:moreLHCcascades}}
\end{figure}

\begin{table}
\caption{Spectrum and branching ratios for $m_{1/2}=500$~GeV,
  $\tan\beta=10$, $m_{H_1}^2=2.7$~TeV$^2$, $m_{H_2}^2=0$.  As the
  first and second generation sfermions are practically degenerate,
  only the first generation is given.}\label{tab:buchmuellerP2}
  \centering
\begin{tabular}{c|r|l}
\hline Sparticle & Mass [GeV] & Dominant decay modes \\ \hline $\ti g$
& 1151.8 & $\ti q_L^{}q$ (15\%), \quad $\ti q_R^{}q$ (37\%), \quad
$\ti b_{1,2}$ (19\%), \quad $\st_1t$ (29\%)\\ $\ti u_L^{}$, $\ti
d_L^{}$ & 1054.0, 1062.0 & $\nt_2\,q$ (32\%),\quad $\ti\x^\pm_1q'$
($\sim$60\%) \\ $\ti u_R^{}$, $\ti d_R^{}$ & 971.8, 1029.2 &
$\nt_1\,q$ (99\%)\\ $\st_1$ & 766.3 & $\nt_1\,t$ (30\%),\quad
$\ti\x^+_1b$ (33\%)\\ $\nt_4$ & 617.9 & $\ti\x^\pm_1W^\mp$
(46\%),\quad $\nt_2h$ (19\%)\\ $\ti\x^\pm_2$ & 614.6 & $\nt_2\,W^\pm$
(26\%),\quad $\ti\x^\pm_1Z$ (22\%)\\ $\nt_3$ & 604.8 &
$\ti\x^\pm_1W^\mp$ (56\%),\quad $\nt_2Z$ (26\%)\\ $\ti e_R$ & 418.3 &
$\nt_1e$ (100\%)\\
   $\stau_2$ & 398.8 & $\nt_1\tau$ (82\%)\\ $\ti\x^\pm_1$ & 387.4 &
   $\ti e_L^\pm\nu_e$ (15\%),\quad $\ti\nu_e e^\pm$ (17\%),\quad
   $\stau_1^\pm\nu_\tau$ (18\%),\quad $\snu\tau^\pm$ (19\%)\\ $\nt_2$
   & 381.3 & $\stau_1^\pm\tau^\mp$ (19\%), \quad $\ti e^\pm_L e^\mp$
   (16\%), \quad $\ti\nu_{e}\nu_{e}$ (15\%)\\ $\ti e_L$ & 206.5 &
   $\nt_1e$ (100\%)\\ $\nt_1$ & 203.4 & $\stau_1^\pm\tau^\mp$ (33\%),
   \quad $\snu\nu_\tau$ (62\%)\\ $\ti\nu_e$ & 198.5 &
   $\snu\nu_e\bar\nu_\tau$ (94\%)\\ $\stau_1$ & 182.3 & $\snu l\nu$
   (32\%),\quad $\snu q\bar q'$ (68\%), \quad $\Gamma=2\times 10^{-8}$
   GeV\\ $\snu$ & 176.1 & $\gravitino\nu_\tau$, \quad
   $\Omega^\text{th}_{\ti\nu}h^2= 7.2\times 10^{-3}$ \\ \hline
\end{tabular}
\end{table}

So far we have assumed $\mnt{2}>m_{\ti l_L}>\mnt{1}$. However, in some
parts of the parameter space the left sleptons can be lighter than the
$\nt_1$, c.f.\ regions C in \fig{snuLSP_region_tb10}.  In this case,
the long decay chains of the type of \fig{LHCcascades}\,(c,\,d,\,e)
obviously do not occur. Instead, we have $\nt_{1,2}\to l^\pm\ti
l^\mp_L$, $\nu_l\ti\nu_l$ and $\ti\x^\pm_1\to \nu\ti l^\pm_L$,
$l^\pm\ti\nu_l$ with $l=(e,\,\mu)$ in addition to the decays into
$\stau_1$ or $\snu$.  These are followed by 3-body decays of the
sleptons: $\ti l_L^\pm\to l^\pm\nu_\tau\snu$, $\nu_l\tau^\pm\snu$ and
$\ti\nu_l\to \nu_l\nu_\tau\snu$, $l^\pm\tau^\mp\snu$.  Some of the
resulting squark decay chains are depicted in \fig{moreLHCcascades}.
A concrete example is realised by taking the parameter point of
\tab{buchmuellerP2} and lowering $m_{1/2}$ to $m_{1/2}=450$~GeV.  The
masses and branching ratios for this case, together with the slepton
decay widths, are given in \tab{P2m12low}.

\begin{table}
\caption{Spectrum and branching ratios for $m_{1/2}=450$~GeV,
  $\tan\beta=10$, $m_{H_1}^2=2.7$~TeV$^2$, $m_{H_2}^2=0$.  As the
  first and second generation sfermions are practically degenerate,
  only the first generation is given.}\label{tab:P2m12low} \centering
\begin{tabular}{c|r|l}
\hline Sparticle & Mass [GeV] & Dominant decay modes \\ \hline $\ti g$
& 1046.1 & $\ti q_L^{}q$ (14\%), \quad $\ti q_R^{}q$ (39\%), \quad
$\ti b_{1,2}$ (18\%), \quad $\st_1t$ (28\%)\\ $\ti u_L^{}$, $\ti
d_L^{}$ & 960.7, 967.6 & $\nt_2\,q$ (32\%),\quad $\ti\x^\pm_1q'$
($\sim$60\%) \\ $\ti u_R^{}$, $\ti d_R^{}$ & 874.9, 940.8 & $\nt_1\,q$
(99\%)\\ $\st_1$ & 685.9 & $\nt_1\,t$ (29\%),\quad $\ti\x^+_1b$
(36\%)\\ $\nt_4$ & 560.5 & $\ti\x^\pm_1W^\mp$ (44\%),\quad $\nt_2h$
(17\%)\\ $\ti\x^\pm_2$ & 557.5 & $\nt_2\,W^\pm$ (25\%),\quad
$\ti\x^\pm_1Z$ (21\%)\\ $\nt_3$ & 545.8 & $\ti\x^\pm_1W^\mp$
(56\%),\quad $\nt_2Z$ (25\%)\\ $\ti e_R$ & 411.1 & $\nt_1e$ (100\%)\\
   $\stau_2$ & 391.2 & $\nt_1\tau$ (83\%)\\ $\ti\x^\pm_1$ & 345.3 &
   $\ti e_L^\pm\nu_e$ (15\%),\quad $\ti\nu_e e^\pm$ (16\%),\quad
   $\stau_1^\pm\nu_\tau$ (18\%),\quad $\snu\tau^\pm$ (19\%)\\ $\nt_2$
   & 339.5 & $\stau_1^\pm\tau^\mp$ (20\%), \quad $\ti e^\pm_L e^\mp$
   (16\%), \quad $\ti\nu_{e}\nu_{e}$ (15\%)\\ $\nt_1$ & 181.4 & $\ti
   e^\pm e^\mp$ (8\%), \quad $\stau_1^\pm\tau^\mp$ (25\%), \quad
   $\snu\nu_\tau$ (32\%)\\ $\ti e_L$ & 142.7 & $\snu\tau\nu_e$
   ($\sim$100\%), \quad $\Gamma= 6\times 10^{-7}$ GeV\\ $\ti\nu_e$ &
   136.5 & $\snu\nu_e\nu_\tau$ (91\%), \quad $\snu e^-\tau^+$ (9\%),
   \quad $\Gamma=4\times 10^{-7}$ GeV\\ $\stau_1$ & 106.0 & $\snu
   l\nu$ (30\%),\quad $\snu q\bar q'$ (70\%), \quad $\Gamma=6\times
   10^{-9}$ GeV\\ $\snu$ & 101.3 & $\gravitino\nu_\tau$, \quad
   $\Omega^\text{th}_{\ti\nu}h^2= 5.5\times 10^{-3}$ \\ \hline
\end{tabular}
\end{table}

A special situation arises for larger $m_{H_2}^2$, as in the right
panels of \figs{snuLSP_region_tb10} and \ref{fig:snuLSP_region_tb20},
in which case the $\mu$ parameter becomes smaller.  Consequently, the
$\nt_{3,4}$ and $\ti\x^\pm_2$ are lighter than in the previous
examples, and the $\nt_{1,2}$ and $\ti\x^\pm_1$ acquire sizable
higgsino components. The $\ti q_{L}^{}$ then decays dominantly into
$\nt_4 q$ and $\ti\x^\pm_{2} q'$, while the $\ti q_{R}^{}$ decays not
only into $\nt_1 q$ but also into $\nt_2 q$. The heavy neutralino and
chargino, $\nt_4$ and $\ti\x^\pm_{2}$, decay further into sleptons,
gauge bosons, or $h^0$ with roughly comparable rates.  This makes this
scenario even more complicated than that of \tab{buchmuellerP2}. The
detection of the heavier neutralino and chargino states through their
decays into sleptons has been studied in \cite{Polesello:2004aq}, and
the use of hadronic neutralino/chargino decays very recently in
\cite{Butterworth:2007ke}.

A comment is in order concerning the detectability of the soft
leptons. For the parameter point of \tab{buchmuellerP2} with
$m_{\stau_1}-m_{\snu}\simeq 6$~GeV, for instance, the mean $p_T$ of
the electrons and muons coming from the $\stau_1\to W^*\snu$
decay is 5.9~GeV at generator level.\footnote{We thank 
Are Raklev for providing the $p_T$ spectrum.} 
Requiring $p_T(e,\mu)>3$~GeV, $5$~GeV, or $10$~GeV
in the offline reconstruction, about 60\%, 40\%, or 17\%,
respectively, of these leptons would pass.  At first glance this 
may appear very challenging for LHC analyses.  Notice, however, 
that the SUSY events can be selected by triggering on the hard
jets/leptons and the $\Emiss$, so that the detection of additional soft 
electrons and/or muons may well be feasible. 
Cuts of $p_T(e)>5$~GeV and $p_T(\mu)>3$~GeV were, for example, 
also used in \cite{CMS:PTDR} for Higgs boson search in the 
$H\to ZZ^{(*)}\to 4l$ channel.  
The situation is of course better for
larger $\snu$--$\stau_1$ mass difference.  Taus and jets coming from
the 3-body $\stau_1$ decays will, however, hardly be observable.

At the ILC~\cite{Aguilar-Saavedra:2001rg,Abe:2001np,Abe:2001gc},
several distinctive features of the $\snu$ NLSP scenario may be
resolved with high accuracy, in particular the large mass splitting
between left and right sleptons with $\msl{L}<\msl{R}$ (although
measuring $\msl{R}$ may require a 1 TeV linear collider).
Selectron-pair production can give $e^+e^-+\Emiss$ or
$e^+e^-\tau^+\tau^-+ 2(f\!\bar f')+\Emiss$, and analogously for smuons
and for $\stau_2$, depending on the mass orderings.  (For
$\mse{L}<\mnt{1}$, however, pair production of $\ti e_L$ leads to
$\tau^+\tau^-+\Emiss$ due to 3-body $\ti e_L$ decays.)  Beam
polarisation, angular distributions and tunable energy can be
exploited to determine the mass, chirality and spin of the sleptons.

Pair production of $\stau_1$ gives $2(f\!\bar f')+\Emiss$. Since the
3-body stau decay proceeds dominantly through an off-shell $W$ boson,
this results in soft jets plus missing energy in half of the cases.
In addition, about 20\% of the $\stau_1\stau_1^*$ events give jets
plus a single charged lepton plus $\Emiss$, and the remaining
$\sim10\%$ lead to $l^\pm l^\mp+\Emiss$ or mixed-flavour events of,
for instance, $e^\pm\mu^\mp+\Emiss$.  On the one hand this certainly
complicates the analysis, on the other hand resolving the various
$l\nu_l$ and $q\bar q'$ modes of the $\stau_1$ decay and estimating
the lifetime allows one to distinguish this scenario from a stau NLSP
which decays into $\tau\gravitino$~\cite{Buchmuller:2004rq,
Hamaguchi:2004df, Feng:2004yi, Martyn:2006as, Ellis:2006vu,
Hamaguchi:2006vu}, $\tau$ axino~\cite{Brandenburg:2005he} or even from
the case of gravitino DM with R-parity
breaking~\cite{Buchmuller:2007ui}.

Chargino production and subsequent decay into lepton and sneutrino
could also provide an efficient way to measure the sneutrino mass, 
as in the case of neutralino LSP studied in~\cite{Freitas:2005et}.

Last but not least, pair-production of $\nt_1$ can lead to visible
events from $\nt_1\to\tau^\pm\stau_1^\mp$ decays, and in the case that
$\mnt{1}>m_{\ti e_L}$ also from $\nt_1\to e^\pm\ti e_L^\mp$,
$\mu^\pm\ti\mu_L^\mp$ decays.  The ISR photon spectrum may give
additional information on the $\nt_1$ and $\snu$ masses.

\section{Conclusions} 

We have considered the case of gravitino LSP and dark matter with a
sneutrino NLSP in the scenario of gaugino-mediated supersymmetry
breaking.  We find viable regions of the parameter space, where the
primordial sneutrino abundance satisfies the BBN constraints. A
general feature of this scenario is a small mass splitting between the
$\stau_1\sim\stau_L$ and the $\snu$, leading to 3-body $\stau_1$
decays into $f\!\bar f'\snu$, dominantly mediated by a virtual
$W$. This can significantly influence the SUSY collider signatures. We
have discussed these signatures depending on the mass ordering of
$\nt_{1,2}$, $\stau_1$ and $\ti e_L$.  In particular, if
$\mnt{1}>\mstau{1}+m_\tau$ (and/or $\mse{L}$), the lightest neutralino
can have visible decays into a charged lepton and slepton.  Moreover,
for $\mnt{1}>\mse{L}$, also selectrons and smuons will only have
3-body decays into the $\snu$. These 3-body decays do, however, not
lead to displaced vertices unless the spectrum is quasi-degenerate.

In general this scenario predicts more soft leptons or jets in the
final states and longer decay chains.  Detailed simulation studies
will be necessary to assess the experimental precisions achievable at
the LHC or ILC in the scenarios discussed here.  This is, however,
beyond the scope of this letter.

\section*{Acknowledgements} 

We would like to thank Wolfgang Adam, Ben Allanach, J\"orn Kersten, 
Giacomo Polesello, Alexander Pukhov and Kai Schmidt-Hoberg,
for useful discussions.

S.K.\ is supported by an APART (Austrian Programme for Advanced
Research and Technology) grant of the Austrian Academy of
Sciences. L.C.\ acknowledges the support of the ``Impuls- und
Vernetzungsfonds'' of the Helmholtz Association, contract number
VH-NG-006.


\end{document}